\documentclass[twocolumn,trackchanges]{aastex701}
\usepackage{graphicx}

\usepackage{booktabs}
\usepackage{array}
\usepackage{rotating}
\usepackage{epstopdf}
\newcommand\nicer{{\it NICER}}
\newcommand\nustar{{\it NuSTAR}}

\newcommand\sax{{\it BeppoSAX}}
\newcommand\chandra{{\it Chandra}}

\newcommand\rxte{{\it RXTE}}
\newcommand\maxi{{\it MAXI}}

\newcommand\inte{{\it INTEGRAL}}

\newcommand\s{{\rm~s}}

\newcommand\kev{{\rm~keV}}

\newcommand\kms{\ifmmode {\rm~km\ s}^{-1} \else ~km s$^{-1}$\fi}

\newcommand\Hunit{\ifmmode {\rm~km\ s}^{-1}\ {\rm Mpc}^{-1}
	\else ~km s$^{-1}$ Mpc$^{-1}$\fi}

\newcommand\ctssec{\ifmmode {\rm~count\ s}^{-1}
	\else ~count s$^{-1}$\fi}

\newcommand\ergsec{\ifmmode {\rm~erg\ s}^{-1}
	\else ~erg s$^{-1}$\fi}

\newcommand\funit{\ifmmode {\rm~erg\ s}^{-1}\;{\rm cm}^{-2}
	\else ~ergs s$^{-1}$ cm$^{-2}$\fi}

\newcommand\phflux{\ifmmode {\rm~photon\ s}^{-1}\;{\rm cm}^{-2}
	\else ~photon s$^{-1}$ cm$^{-2}$\fi}

\newcommand\efluxA{\ifmmode {\rm~erg\ s}^{-1}\;{\rm cm}^{-2}\;{\rm \AA}^{-1}
	\else ~erg s$^{-1}$ cm$^{-2}$ \AA$^{-1}$\fi}

\newcommand\efluxHz{\ifmmode {\rm~erg\ s}^{-1}\;{\rm cm}^{-2}\;{\rm Hz}^{-1}
	\else ~erg s$^{-1}$ cm$^{-2}$ Hz$^{-1}$\fi}

\newcommand\cc{\ifmmode {\rm~cm}^{-3} \else cm$^{-3}$\fi}

\newcommand\FWHM{\ifmmode {\rm~FWHM} \else ${\rm~FWHM}$\fi}

\newcommand\Msun{\ifmmode M_{\odot} \else $M_{\odot}$\fi}

\newcommand\Lsun{\ifmmode L_{\odot} \else $L_{\odot}$\fi}

\newcommand\hbeta{\ifmmode {\rm H}\beta \else H$\beta$\fi}

\newcommand\Kalpha{\ifmmode {\rm K}\alpha \else K$\alpha$\fi}

\newcommand\nh{\ifmmode N_{\rm H} \else N$_{\rm H}$\fi}

\begin{document}

\title{The first comprehensive spectral and timing study of the ultra-compact X-ray binary 4U~1812-12 with \nicer{} and \nustar{}}

\author[orcid=0000-0003-4950-9134
,gname=Swarnendu,sname='Jana']{Swarnendu Jana}
\affiliation{{Department of physics, Visva-Bharati, Santiniketan, West Bengal, 731235, India}}
\email{swarnendujana.rs.phys@visva-bharati.ac.in}
\author[orcid=0000-0002-2565-1219,gname=Aditya S., sname='Mondal']{Aditya S. Mondal} 
\altaffiliation{Email:adityas.mondal@visva-bharati.ac.in}
\affiliation{{Department of physics, Visva-Bharati, Santiniketan, West Bengal, 731235, India}}
\email{adityas.mondal@visva-bharati.ac.in}

\author[orcid=0000-0003-3753-3102,gname=Aru, sname='Beri']{Aru Beri}
\affiliation{Indian Institute Of Astrophysics (IIA), Koramangala, Bengaluru, Karnataka, 560034, India}
\email{aru.beri@iiap.res.in}
\author[orcid=0000-0003-1589-2075,gname=Gulab, sname='Dewangan']{Gulab C. Dewangan}
\affiliation{Inter-University Centre for  Astronomy \& Astrophysics (IUCAA), Pune, 411007, India}
\email{gulabd@iucaa.in}

\begin{abstract}
The source 4U~1812-12 is a persistent, weakly variable low-mass X-ray binary containing a neutron star. The source was observed by \nicer{} between 2019 and 2021 and, more recently, by \nustar{} in 2025. During the \nicer{} and \nustar{} observations, the source was detected in a hard spectral state with a bolometric luminosity of $\sim 1.90\times 10^{36}$ ergs s$^{-1}$. Its $3-70$ \kev{} \nustar{} spectrum is characterized by a soft thermal emission from the disc, a hard Comptonized emission from the corona, and its reflection from the accretion disc. The \nustar{} energy spectrum exhibits the clear presence of disc reflection features, fitted using a self-consistent relativistic reflection model {\tt relxill}. Our reflection modeling indicates a moderately ionized accretion disc (log\:$\xi\sim2.72$) extending close to the neutron star surface ($R_{in}\lesssim 1.72\:R_{ISCO}$), and viewed through a small inclination angle ($i\sim 25$ degrees). Assuming that the magnetic field ($B$) truncates the disc, we found $B\lesssim 2.54\times 10^{8}$ G, comparable to the typical values observed for NS LMXBs. The $1.0-9.5$ \kev{} \nicer{} spectra are also characterized by a soft thermal component and a dominant hard Comptonized component. During \nicer{} observations, the disc temperature exhibits a small variation within $\sim 0.69-0.84$ \kev{}. In contrast, the power law photon index, $\Gamma$, exhibits a large variation of $\sim 0.8-1.5$, implying a substantial change in the Comptonized emission. Moreover, \nicer{} timing analysis reveals broadband aperiodic variability with significant QPO-like features at $0.379\pm 0.008$ Hz and $0.724\pm 0.025$ Hz, having fractional rms amplitudes of $2.9\pm 0.6\%$ and $4.1\pm 0.5\%$, respectively.

\end{abstract}


\section{Introduction}
Ultra-compact X-ray binaries are a rare class of interacting binary systems in which a compact object, either a neutron star or a black hole, accretes matter from a degenerate companion, such as a He star or a white dwarf. These X-ray binaries have extremely short orbital periods, typically less than 80 minutes \citep{1982ApJ...254..616R}. Even though a large number of ultra-compact X-ray binaries are predicted to be hosted in our Galaxy \citep{2013A&A...552A..69V}, the current population of these sources is limited to only 14 systems with reported orbital periods \citep{2013ApJ...768..184H, 2018ApJ...858L..13S}. A large number of systems remain undetected due to observational challenges in measuring their orbital periods and to their faintness \citep{2017hsn..book.1499C}. Mass transfer in ultra-compact X-ary binaries occurs through Roche-lobe overflow, forming a small accretion disc around the neutron star \citep{1994A&A...290..133V}. Ultra-compact systems exhibit a persistent behaviour when accreting at lumonisities $\lesssim 10^{36}$ erg s$^{-1}$, while systems with longer orbits have persistent X-ray luminosities of $\sim 10^{37-38}$ erg s$^{-1}$ \citep{2010NewAR..54...87N}.\\

The neutron star low-mass X-ray binary (LMXB) 4U~1812-12 is considered an ultra-compact X-ray binary as it fulfills several distinctive features of this type of X-ray binary. The source was first detected by the Uhuru mission in 1970 \citep{1976ApJ...206L..29F}. It was later observed by several satellites, including \textit{OSO-7}, \textit{Ariel V}, \textit{HEAO-1}, and \textit{EXOSAT} with a consistent low X-ray flux of $\sim (3-6)\times 10^{-10}$ ergs cm$^{-2}$ s$^{-1}$ (\citealt{1995xrbi.nasa..536V}, and references therein). X-ray bursts were first detected from this source in 1982 with Hakucho \citep{1983PASJ...35..531M} and later with BeppoSAX \citep{2000A&A...357..527C}. The detection of type-I X-ray bursts confirms the neutron star nature of the compact object. The source 4U~1812-12 is further classified as an atoll source like most of the bursters, and shows band-limited noise and a $\sim 0.8$ Hz QPO \citep{1999ApJ...514..939W}. The distance of the source is $\sim 4.1$ kpc, as estimated assuming Eddington-limited luminosities of some type-I bursts showing the characteristics of photospheric radius expansion \citep{2000A&A...357..527C}. This distance measurement implies a persistent low X-ray luminosity of $\sim 10^{36}$ erg s$^{-1}$. Its persistent low X-ray luminosity and low optical-to-X-ray flux ratio suggest that it harbors a small accretion disc \citep{1994A&A...290..133V, 2006A&A...446L..17B}. Moreover, its optical spectrum does not exhibit hydrogen spectral features, which suggests a hydrogen-exhausted/hydrogen-poor donor star \citep{2020A&A...644A..63A}. However, recent photometric studies suggest that the source 4U~1812-12 is most likely not an ultracompact X-ray binary but a strong ultra-compact candidate or the progenitor of an ultra-compact X-ray binary, as it shows a tentative $\sim 114$ minutes orbital period \citep{2022ApJ...931L...9A}.\\

A broad band spectrum of the persistent emission and the power density spectrum of the source rapid X-ray variability have been revealed during simultaneous observations with \sax{} and \rxte{} on April 20, 2000 \citep{2003A&A...400..643B}. The source was observed by \sax{} in a hard spectral state with a bolometric luminosity of $\sim 2\times 10^{36}$ erg s$^{-1}$. Its broad-band energy spectrum is characterized by the presence of a hard X-ray tail extending above $\sim 100$ kev{} \citep{2003A&A...400..643B}. Its power density spectrum is characterized by the presence of a $\sim 0.7$ Hz low frequency QPO together with three broad noise components \citep{2003A&A...400..643B}. \chandra{} observed the source on June 14, 2000, and the persistent emission is characterized by a $2-10$ \kev{} flux of $\sim 4\times 10^{-10}$ ergs cm$^{-2}$ s$^{-1}$ \citep{2003ApJ...596.1220W}. \rxte{} again observed the source in June and July 2001, and the source was in the hard state with a $2-10$ \kev{} flux of $\sim 3.8\times 10^{-10}$ ergs cm$^{-2}$ s$^{-1}$ \citep{2005ApJ...626.1020M}. The source was further monitored with the \inte{} observatory during 2003-2004. During this observation, the source was observed in a hard spectral state with a $1-200$ keV luminosity of $2\times 10^{36}$ erg s$^{-1}$ \citep{2006A&A...448..335T}. Later, \nicer{} observed the source on many occasions between 2019 and 2021, and \nustar{} observed the source recently in 2025.\\

In the present work, we have carried out a detailed spectral and timing analysis of the source 4U~1812-12 using recent \nustar{} and \nicer{} observations. For the first time, we have presented a comprehensive spectral analysis of the source to precisely probe the accretion geometry. From the \nustar{} spectrum, we detected X-ray reflection of the hard Comptonized emission from the accretion disc, not detected earlier by any observation, and applied a self-consistent relativistic reflection model to investigate this further. Moreover, we explored the presence of QPO-like features by constructing the power density spectrum (PDS) of many \nicer{} observations performed between 2019 and 2021. This paper is structured as follows. In Section 2, we describe the observations and data reduction techniques used in this study, and in Section 3, we discuss the timing analysis, including characteristics of the source's light curves and power density spectrum. Section 4 describes the spectral analysis techniques adopted for the continuum and reflection emission. Finally, we discuss the observational findings and their implications in Section 5.\\

\begin{table*}[ht]
\centering
\caption{Summary of the NuSTAR observation of 4U~1812-12.}
\renewcommand{\arraystretch}{1.2}
\begin{tabular}{lccccc}
\toprule
\textbf{ObsID} &
\textbf{ShortID} &
\textbf{Observed Date} &
\textbf{Instrument} &
\textbf{Exposure} &
\textbf{Count Rate} \\
&
&
&
&
\textbf{(ks)} &
\textbf{(cts s$^{-1}$)} \\
\midrule
31101029002 & Obs 1 & 2025-09-19 & FPMA/B & $\sim$ 102 & 12.0 \\
\bottomrule
\end{tabular}
\label{nustar_obs}
\end{table*}
\begin{table*}
\centering
\caption{Summary of the NICER observations of 4U~1812-12 that are included in this work.}
\footnotesize
\renewcommand{\arraystretch}{1.5}
\begin{tabular}{lcccccc}
\toprule
\textbf{ObsID} &
\textbf{ShortID} &
\textbf{Observed Date} &
\textbf{Exposure} &
\textbf{Count Rate} &
\textbf{Flaring} &
\textbf{Absorption Dips} \\
&
&
&
\textbf{(ks)} &
\textbf{(cts/ s)} &
\textbf{(Yes/No)} &
\textbf{(Yes/No)} \\
\midrule
3580010103 & Obs 2  & 2020-04-25 & $\sim$ 25 & 53.85 & Yes & No \\
3580010203 & Obs 3  & 2020-07-11 & $\sim$ 21 & 45.63 & No & No \\
3580010202 & Obs 4  & 2020-07-10 & $\sim$ 21 & 43.87 & No & No \\
3580010104 & Obs 5  & 2020-04-26 & $\sim$ 21 & 53.85 & No & Yes \\
2560010104 & Obs 6  & 2019-03-09 & $\sim$ 19 & 33.73 & No & No \\
2560010103 & Obs 7  & 2019-03-07 & $\sim$ 18 & 33.32 & No & No \\
2560010204 & Obs 8  & 2019-06-27 & $\sim$ 17 & 40.40 & No & No \\
2560010203 & Obs 9  & 2019-06-26 & $\sim$ 17 & 40.60 & Yes & No \\
2560010102 & Obs 10 & 2019-03-07 & $\sim$ 16 & 32.83 & No & No \\
3580010204 & Obs 11 & 2020-07-12 & $\sim$ 15 & 46.89 & No & No \\
3580010102 & Obs 12 & 2020-04-24 & $\sim$ 15 & 52.46 & No & No \\
4202290101 & Obs 13 & 2021-08-09 & $\sim$ 14 & 49.45 & No & No \\
4202290102 & Obs 14 & 2021-08-10 & $\sim$ 14 & 50.44 & No & No \\
3580010205 & Obs 15 & 2020-07-13 & $\sim$ 10 & 47.51 & No & No \\
3580010105 & Obs 16 & 2020-04-26 & $\sim$ 7 & 54.46 & No & No \\
3580010201 & Obs 17 & 2020-07-09 & $\sim$ 7 & 43.72 & No & No \\
2560010202 & Obs 18 & 2019-06-25 & $\sim$ 6 & 37.37 & No & No \\
2560010205 & Obs 19 & 2019-06-27 & $\sim$ 5 & 41.89 & No & No \\
3580010307 & Obs 20 & 2020-08-21 & $\sim$ 5 & 41.75 & No & No \\
2560010101 & Obs 21 & 2019-03-06 & $\sim$ 4 & 33.14 & No & No \\
3580010308 & Obs 22 & 2020-08-22 & $\sim$ 4 & 41.04 & No & No \\
4202290103 & Obs 23 & 2021-08-11 & $\sim$ 4 & 50.61 & No & No \\
3580010315 & Obs 24 & 2020-09-04 & $\sim$ 3 & 45.36 & No & No \\
4202290104 & Obs 25 & 2021-09-07 & $\sim$ 3 & 55.35 & No & No \\
3580010310 & Obs 26 & 2020-08-25 & $\sim$ 2 & 42.61 & No & No \\
\bottomrule
\end{tabular}
\label{nicer_obs}
\end{table*}

\begin{figure*}
\centering
\includegraphics[scale=0.60]{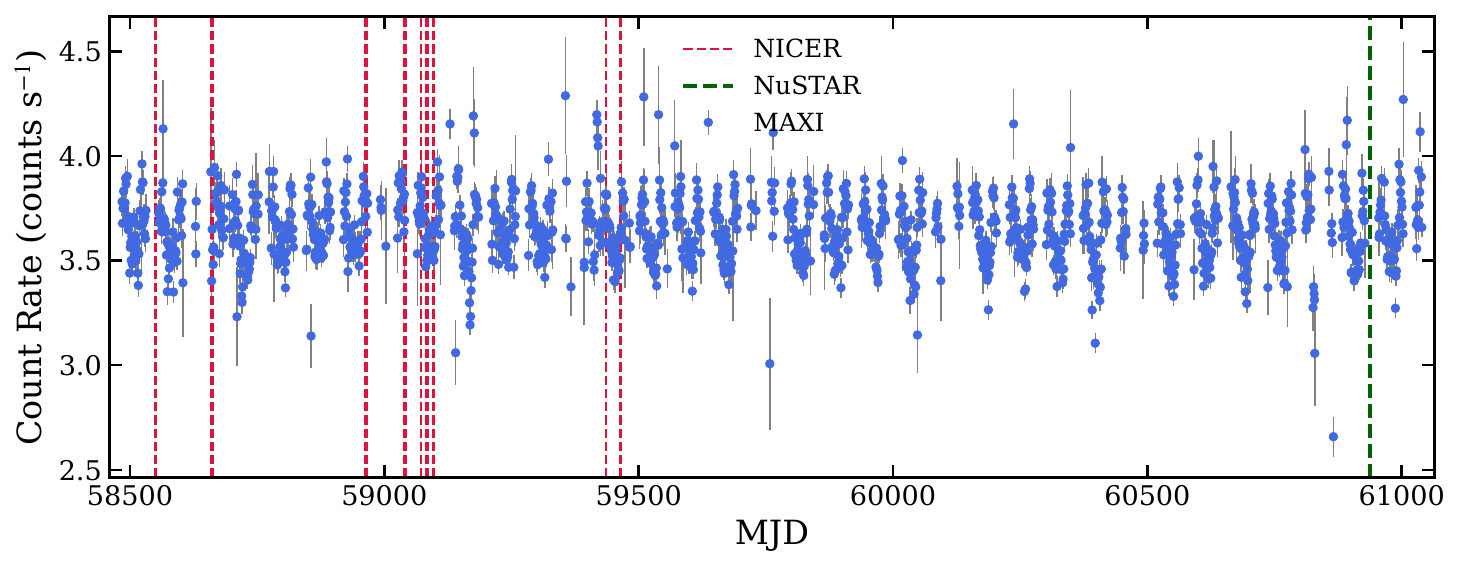}
\includegraphics[scale=0.32]{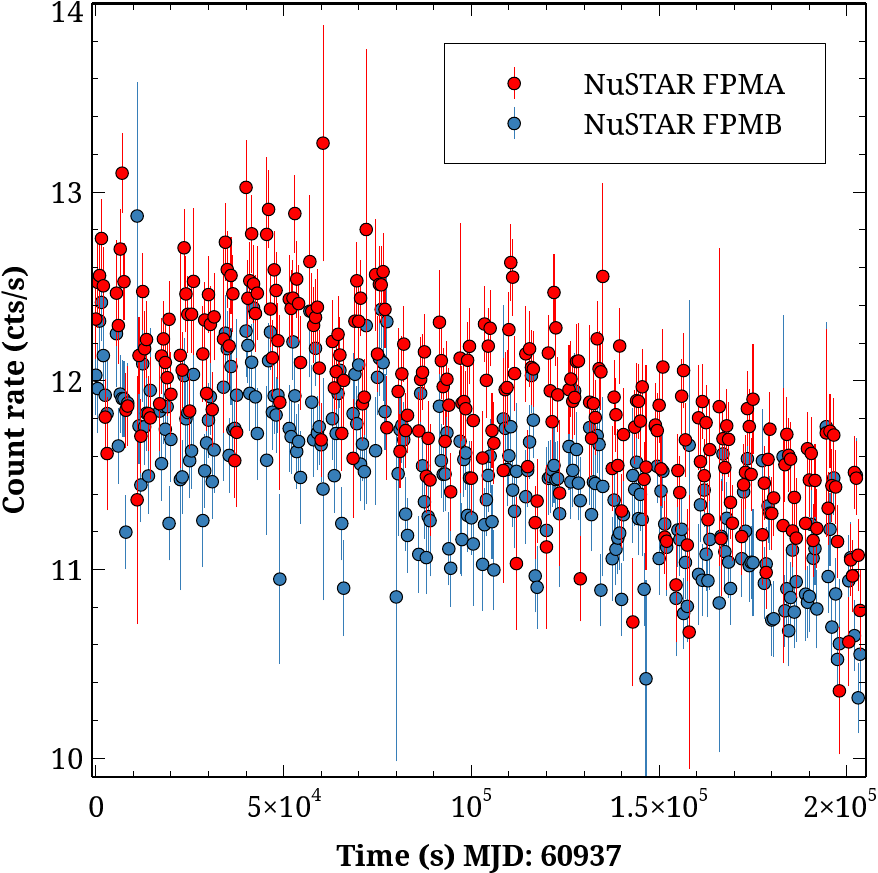}
\includegraphics[scale=0.35]{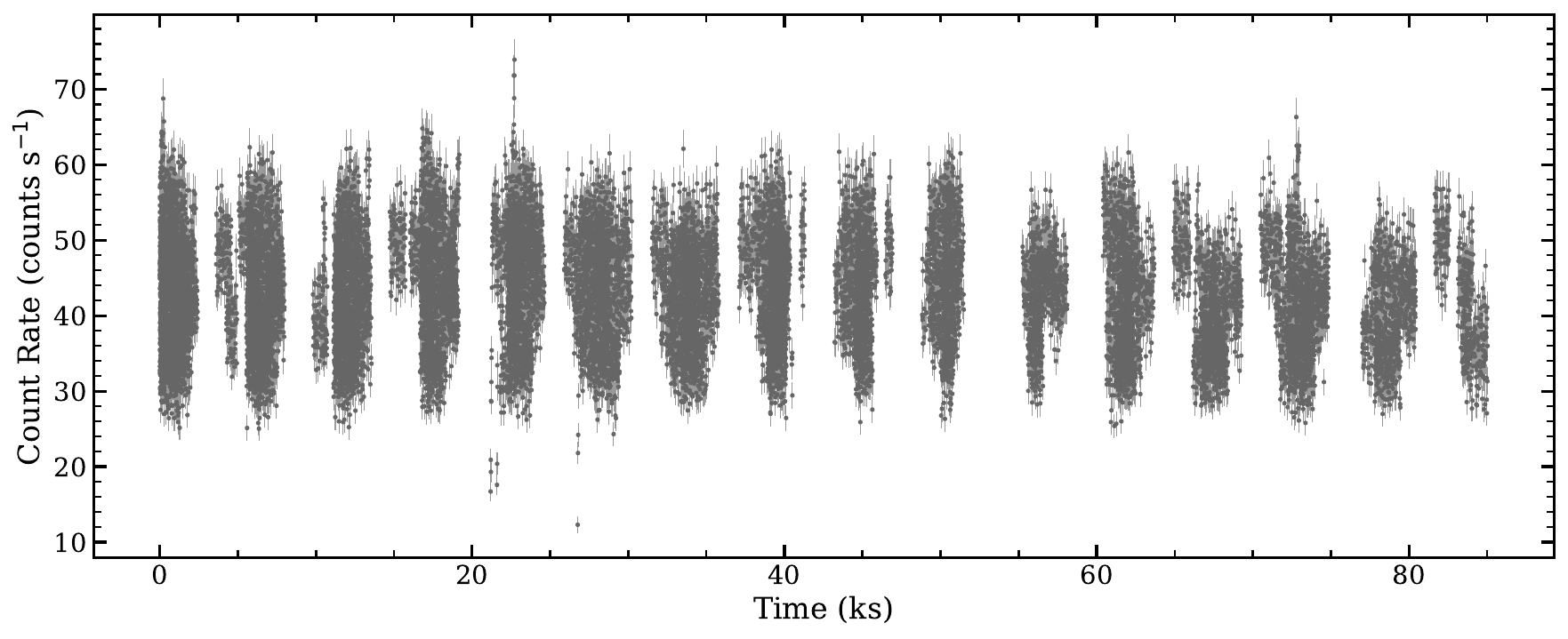}
\includegraphics[scale=0.32]{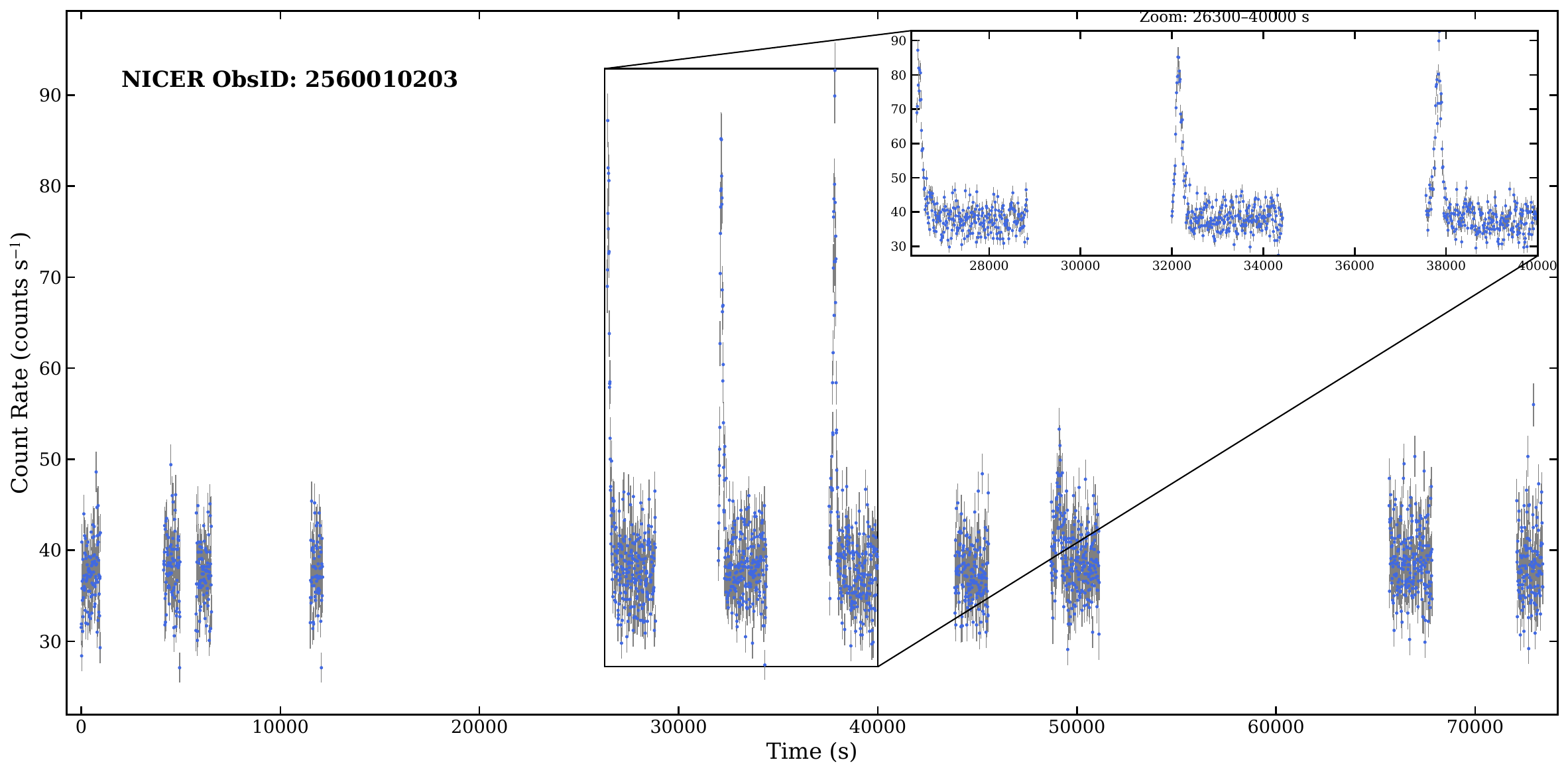}
\caption{Top panel: It shows the \maxi{}/GSC long-term light curve of the source 4U~1812-12 in the $2-20$ \kev{} energy band with a bin size of 1 day. In this light curve, all the \nicer{} and \nustar{} observations, considered for the present analysis, are marked by vertical lines. Middle: (left) It shows the $3-79\kev{}$ \nustar{} FPMA and FPMB light curves with a time bin size of 500 s. (right) It shows the \nicer{} light curve of the source, with a bin size of 10 s, after combining all observations listed in Table~\ref{nicer_obs}, except those which exhibit flare and dips. Bottom panel: It shows the \nicer{} light curve for obs9 (ID: 2560010203), which exhibits regular flaring activity, shown in the inset of this plot.} 
\label{Fig1}
\end{figure*}

\section{observation and data reduction}
The Nuclear Spectroscopic Telescope Array (\nustar{}) is a hard X-ray (3–79 keV) focusing telescope, consisting of two identical, co-aligned focal plane module (FPM) telescopes, A and B. Its effective area peaks at $\sim 900$ cm$^{2}$ (adding up the two modules) around 10 keV, and its energy resolution at 10 keV is 400 eV \citep{2013ApJ...770..103H}. The source 4U~1812-12 was observed with the \nustar{} on 2025 September 19 (Obs ID: 31101029002), as marked by the green vertical line in the Figure~\ref{Fig1}. The observation has a total effective exposure of $\sim 100 $ ks per FPM (see Table~\ref{nustar_obs} for details). The Neutron star Interior Composition Explorer (\nicer{}) X-ray timing instrument (XTI) observed the source on many occasions between 2019 and 2021 \citep{2016SPIE.9905E..1HG}. Among them, we initially considered 25 \nicer{} observations with longer exposure times, preferably above 2 ks. From them, we initially ignored some of the observations for which either flare/X-ray bursts or absorption dips are detected, as listed in the Table~\ref{nicer_obs}. The remaining \nicer{} observations were included in our persistent emission analysis to investigate variation in different system parameters across the \nicer{} observations. We separately considered the \nicer{} observations for which flaring activity is observed.\\

We processed the \nustar{} data using the data analysis Software {\tt NuSTARDAS v2.1.5}, which is distributed with {\tt HEASOFT v6.36}, and used the latest calibration files {\tt v20260421} available during the analysis. We generated cleaned event files using the standard screening criteria with the task {\tt nupipeline v0.4.12}. We extracted a circular source region of radius $100^{''}$ using {\tt SAOImageDS9}, centered on the source position, to study the source spectrum. Background events were extracted using an equivalent radius but from the source-free region for each detector. Spectra, light curves, and response files were produced using the FTOOL {\tt nuproducts} from the FPMA and FPMB. The spectra from both the detectors were grouped, imposing a minimum of $100$ counts per bin. We fitted the spectra in the energy range $3-70$ \kev{} as the background emission dominates above $ 70$ \kev{}.  \\

All \nicer{} data were reduced using the tool {\tt NICERDAS} version 2025-09-05$\_$V015 and latest {\tt CALDB} version {\tt xti20240206}. Cleaned event files were generated from the \nicer{} data using the task \textit{nicerl2}. We created the source spectra and responses from the cleaned event files using the task \textit{nicerl3-spect}. In addition, the light curves of the sources were generated using the task \textit{nicerl3-lc}. We estimated the background using the \textit{scorpeon} background model described by \citet{2022AJ....163..130R}. We use the \nicer{} spectral data in the $1.0-9.5$ \kev{} energy range to avoid several well-studied instrumental and interstellar medium related features beyond this range. 

\section{Timing Analysis}
\subsection{Light Curve}
The top panel of Figure~\ref{Fig1} shows the \maxi{}/GSC light curve of the source 4U~1812-12 in the $2-20$ keV energy band with a bin size of 1 day, where all \nicer{} and \nustar{} observations, considered for the present analysis, are shown by vertical lines. The middle-left panel of Figure~\ref{Fig1} shows the $3-79\kev{}$ \nustar{} FPMA and FPMB light curves with a 500-s time bin. The average count rate during the \nustar{} observation is found to be $\sim 11-13$ counts s$^{-1}$, indicating a persistent emission during \nustar{} observation. The middle-right panel of Figure~\ref{Fig1} shows the \nicer{} light curve of the source, with a bin size of 10 s, after combining all observations listed in Table~\ref{nicer_obs}, except those which exhibit flare and dips. Within the \nicer{} observations, the count rate fluctuates around $\sim 40-50$ counts s$^{-1}$. There is no obvious increase or decrease in the average count rate throughout the \nicer{} observations, suggesting that the source remained in the same spectral or accretion stage. However, occasional flares and absorption dips have been observed in some time intervals, which may be due to intrinsic accretion variability and partial obscuration by disc material. In our persistent spectral analysis, we have ignored those observations that exhibit flares and absorption dips. \\

We note that two \nicer{} observations (obs2 and obs9) show an increase in intensity lasting $\sim 200-300$ seconds in the light curves. For obs2, it was observed once with a maximum count rate of $\sim 120$ counts s$^{-1}$ , whereas for obs9, it occurred three times with a count rate of $\sim 80-90$ counts s$^{-1}$, all at the beginning of the \nicer{} orbit (see bottom panel of Figure~\ref{Fig1}). This behavior is attributed to the flaring activity, as their morphologies differ markedly from those of typical type-I bursts. However, for obs2, the decay of the high-intensity phase resembles that of type-I bursts. Still, it cannot be conclusively identified as a type-I burst, as its rising phase is missing and likely falls within the \nicer{} data gap. Because of this uncertainty, we did not take this high-intensity regime into our analysis. For obs9, we generated GTI files for the flaring regions (denoted as flares 1, 2, and 3) and extracted spectra from these regions for further analysis.

\begin{table*}[!t]
\centering
\caption{Continuum fitting parameters obtained from the \nustar{} observation of the source 4U~1812$-$12 in the $3-70$ keV energy range.}
\setlength{\tabcolsep}{3.5pt}
\renewcommand{\arraystretch}{1.5}

\begin{tabular}{ccccc}
\hline
\textbf{Model Component} & \textbf{Parameter} & \textbf{Model 1} & \textbf{Model 2} & \textbf{Model 3} \\
\hline

Constant & FPMB (with FPMA) & $1.005\pm0.001$ & $1.006\pm0.001$ & $1.006\pm0.001$ \\
TBabs & $N_{\rm H}$ ($10^{22}$ cm$^{-2}$) & 0.63 (fixed) & 0.63 (fixed) & 0.63 (fixed) \\
Diskbb & $kT_{\rm in}$ (keV) & $0.95\pm0.03$ & $0.74\pm0.02$ & $0.81\pm0.04$ \\
& Norm & $5\pm1$ & $14^{+4}_{-2}$ & $7^{+2}_{-1}$ \\
CutoffPL & $\Gamma$ & $1.58\pm0.01$ & -- & -- \\
& $E_{\rm cut}$ (keV) & $76\pm4$ & -- & -- \\
& Norm ($\times10^{-2}$) & $6.13\pm0.16$ & -- & -- \\
nthComp & $\Gamma$ & -- & $1.79\pm0.01$ & $1.79\pm 0.01$ \\
& $kT_e$ (keV) & -- & $22\pm1$ & $22\pm 1$ \\
& $kT_{\rm bb}$ (keV) & -- & $=kT_{\rm in}$ & $\leq 0.90$ \\
& Norm ($\times10^{-2}$) & -- & $4.49\pm0.26$ & $6.64\pm1.6$ \\
& Input Type & -- & 1 (fixed) & 0 (fixed) \\
\hline
 & $\chi^2$/dof & $1608/1506$ & $1637/1506$ & $1633/1505$ \\
\hline
\end{tabular}
\label{continuum}
\end{table*}

\subsection{Power density spectrum}
\begin{figure*}
\centering
\includegraphics[scale=0.42, angle=270]{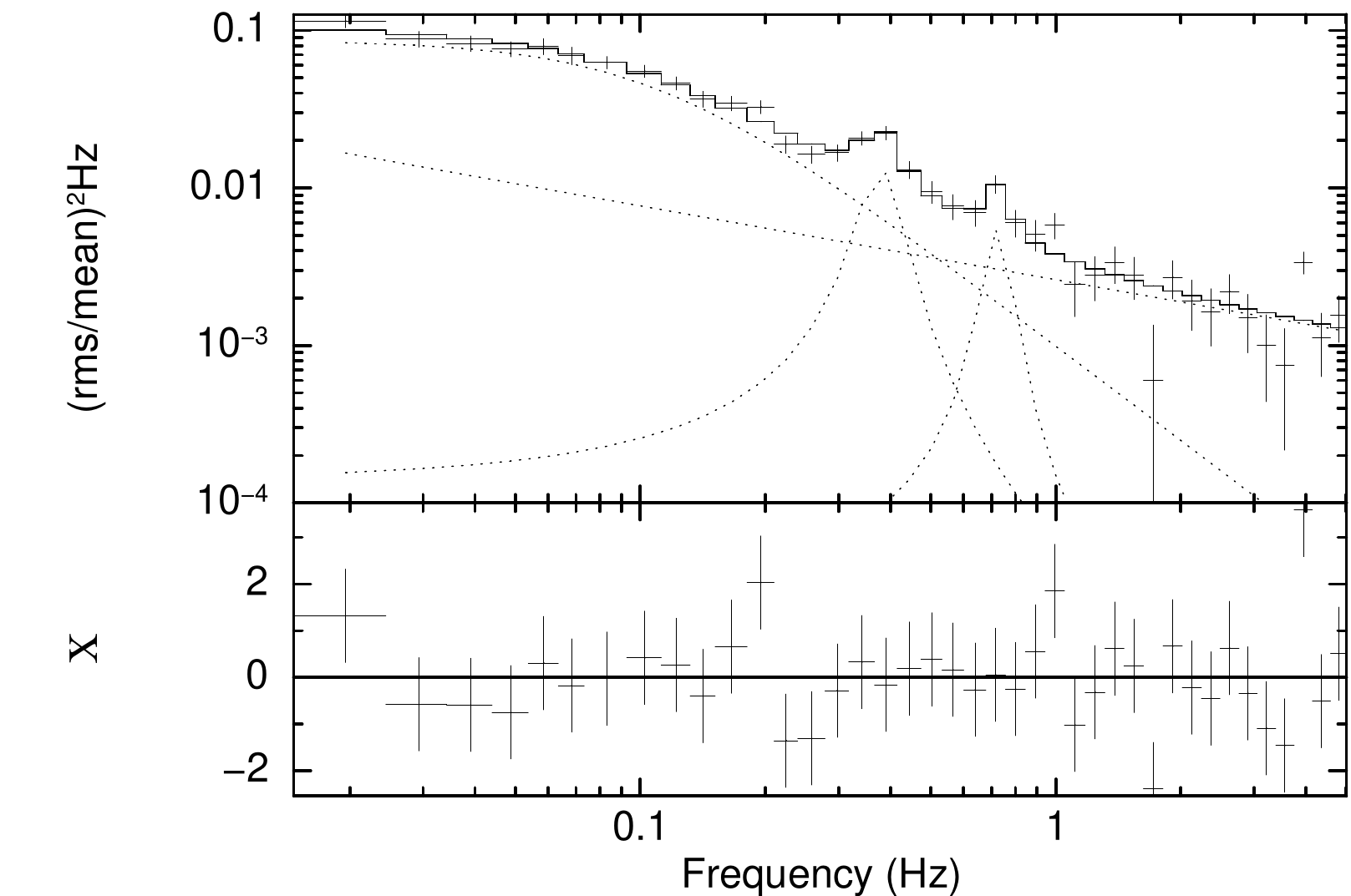}
\caption{The power density spectrum of the source 4U~1812-12 using the \nicer{} obs4 listed in Table~\ref{nicer_obs} (Obs ID: 3580010202). It covers a frequency range of approximately $0.01-5$ Hz and exhibits two narrow Lorentzian components centered at $\sim 0.38$ Hz and $\sim 0.72$ Hz. } 
\label{Fig1_1}
\end{figure*}

The \nicer{} power density spectra (PDS) were generated using the {\tt Heasoft} task {\tt POWSPEC} from a $0.1$-s binned light curve. The PDS were computed from $102.4$-s segments and averaged over the entire observation, covering a frequency range of approximately $0.01-5$ Hz. Most observations are dominated by broadband aperiodic variability and do not show any statistically significant narrow timing features. However, in obs4 (ID: 3580010202), the PDS exhibits two narrow, distinct peaks, as shown in the Figure~\ref{Fig1_1}. To model the PDS, we adopted the standard {\tt multi-Lorentzian} approach (e.g., \citealt{2000MNRAS.318..361N, 2000A&A...355..271B}). The resulting spectrum was fitted in {\tt XSPEC} using a model consisting of three Lorentzian components and a power-law continuum. The fit provided an acceptable description of the data with $\chi^2/dof=41.1/40$. One of the Lorentzian components was fixed at zero centroid frequency to account for the zero-centered broad-band noise, while the centroid frequencies of the remaining two components are allowed to vary freely. The dominant variability is represented by a broad zero-centered Lorentzian with a width of $0.215\pm 0.016$ Hz. The resulting centroid frequencies of the Lorentzian components are $0.379\pm 0.008$ Hz and $0.724\pm 0.025$ Hz. The two narrow components correspond to quality factors of $Q\sim 5$ and $Q\sim 9$, indicating enhanced variability at discrete frequencies superposed on the broad-band noise continuum. A weak power-law component with index $0.47\pm 0.18$ accounts for the residual low-frequency variability. The PDS and best-fitting model are shown in Figure~\ref{Fig1_1}. We further computed the fractional root mean square (rms) amplitude associated with the two {\tt Lorentzian} components that describe the quasi-periodic oscillation (QPO)-like feature in the PDS. The fundamental and (sub)harmonic features have fractional rms amplitudes of $2.9\pm 0.6\%$ and $4.1\pm 0.5\%$, respectively.

\section{spectral analysis}
The \nicer{} and \nustar{} spectra are analyzed with the X-ray spectral package {\tt XSPEC} $v12.15.0$ \citep{1996ASPC..101...17A}. We model the \nustar{} FPMA and FPMB spectra simultaneously over the $3$ to $70$\kev{} energy band, as there is a background dominance above $70$ \kev{}. To account for cross-instrument calibration between FPMA and FPMB, a constant multiplication factor {\tt constant} was included in our modeling. The value of {\tt constant} for \nustar{}/FPMA was fixed to $1$, allowing it to vary for the FPMB. We fit the \nicer{} spectrum in the energy band $1.0-9.5$ keV for all observations. We used the {\tt TBabs} model to account for galactic absorption along the line of sight with the {\tt wilm} abundances \citep{2000ApJ...542..914W} and the {\tt vern} \citep{1996ApJ...465..487V} photoelectric cross-section. Spectral uncertainties are quoted at the $1 \sigma$ confidence level.

\subsection{Continuum emission}
\subsubsection{NuSTAR}
\begin{figure*}
\centering
\includegraphics[width=0.48\textwidth, scale=0.30]{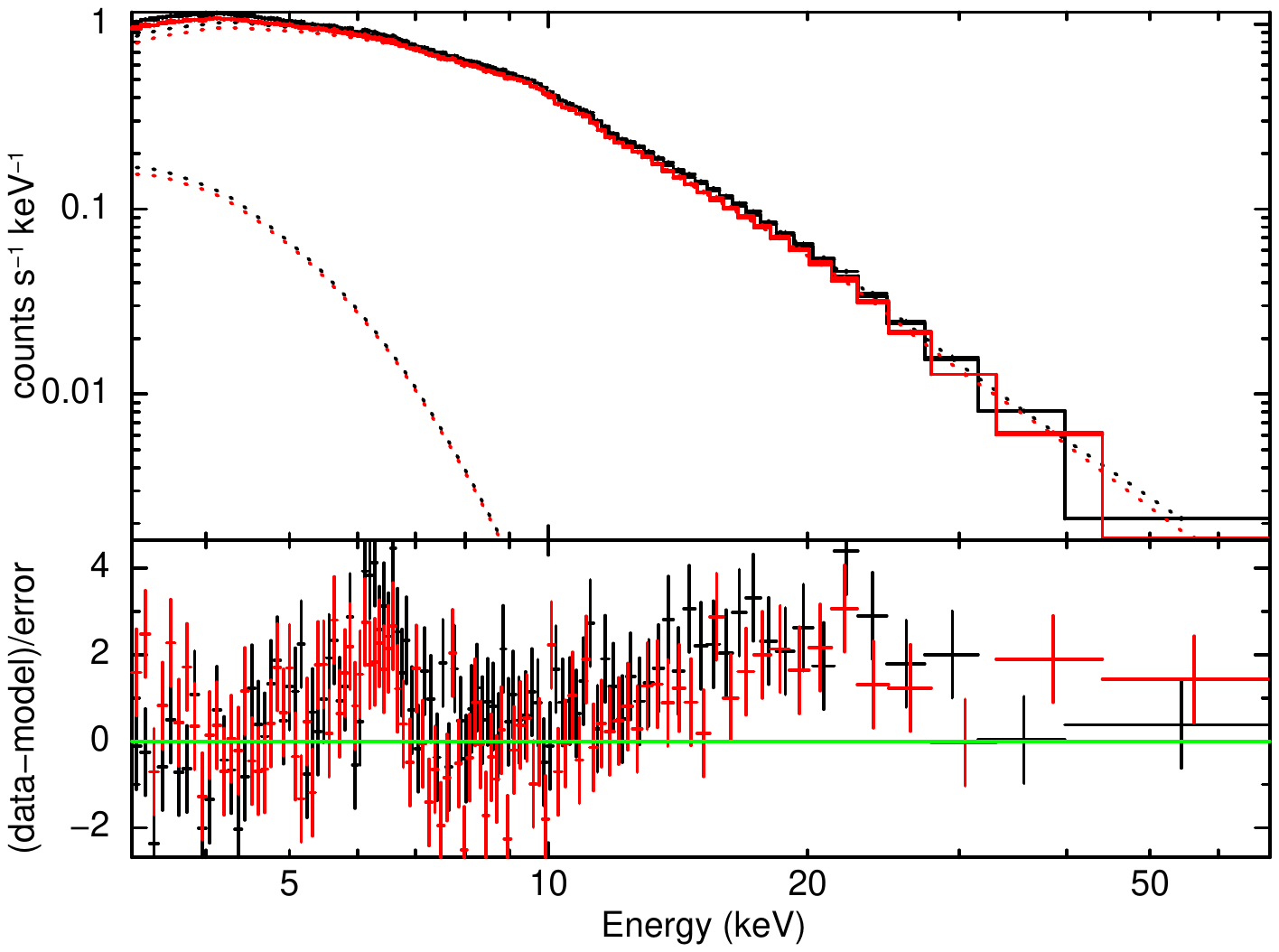}
\includegraphics[width=0.48\textwidth, scale=0.30]{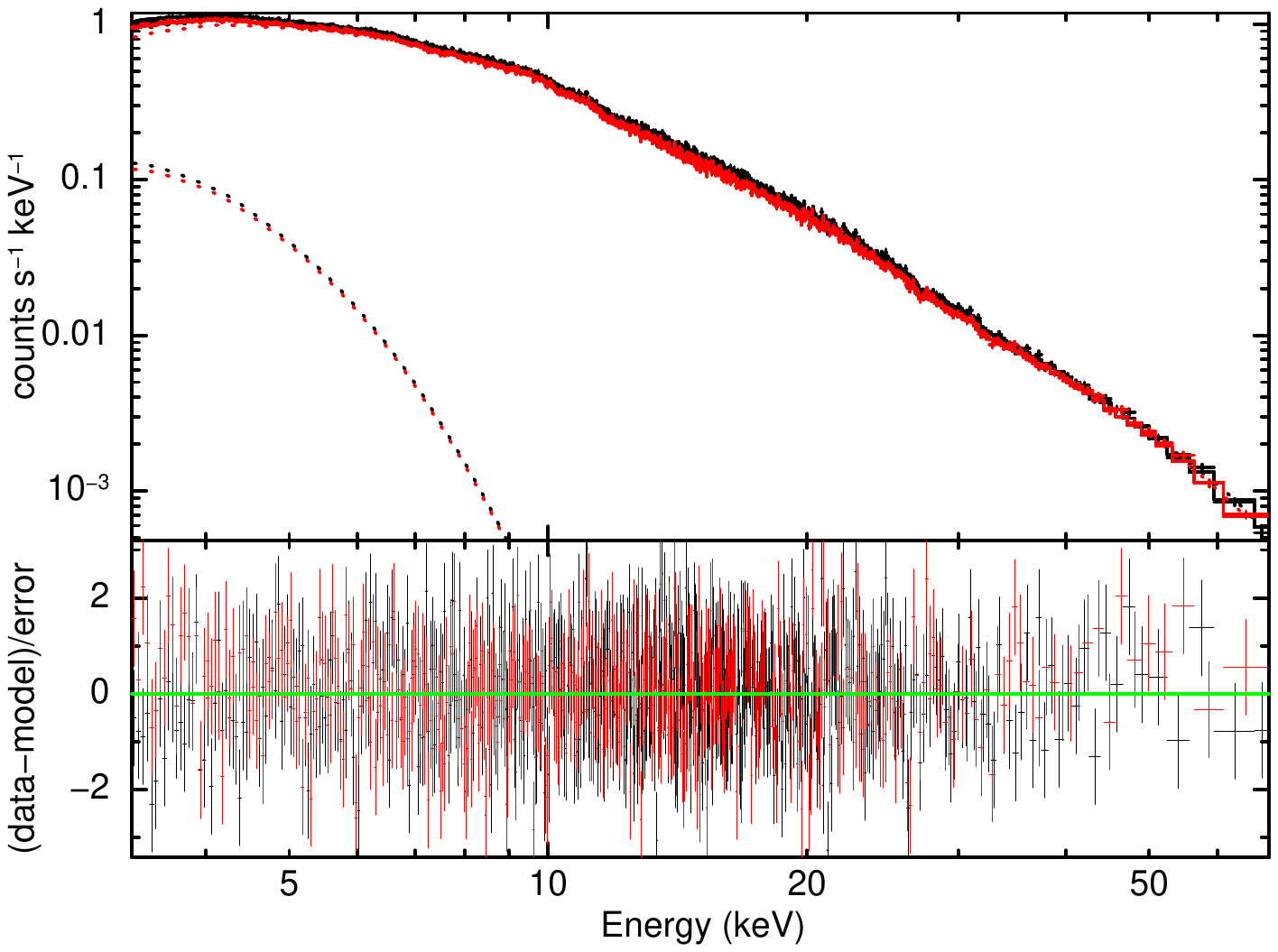}
\caption{Left: The \nustar{} FPMA (black) and FPMB (red) energy spectrum of the source 4U~1812-12 in the energy band $3-70$ keV. The unfolded spectral data for the continuum model {\tt const*TBabs*(diskbb+powerlaw)} is shown. The reflection features are evident from the spectrum.  Left: The unfolded spectral data for the best-fit model {\tt const*TBabs*(diskbb+relxill)} is shown. The lower panel of both plots shows the ratio of the data to the model in units of $\sigma$. The data have been rebinned for visual purposes.}
\label{Fig2}
\end{figure*}

Initially, we tried to fit the \nustar{} spectrum using the absorbed cut-off power-law model ({\tt cutoffpl} in {\tt XSPEC}). However, it failed to model the continuum emission ($\chi^2/dof=2270/1508$), leaving large residuals in the soft-energy range. We therefore added a soft thermal component ({\tt diskbb} in {\tt XSPEC}), which significantly improved the fit, yielding a $\chi^2/dof$ value of $1608/1506$ (the $f-$test rejection probability is extremely low $p\sim 10^{-112}$). Therefore, the continuum emission is well described by the combination of the model {\tt diskbb} and {\tt cutoffpl}, i.e., Model 1: {\tt const*tbabs*(diskbb+cutoffpl)}, where the {\tt diskbb} and {\tt cutoffpl} components are used to account for the emission from the disc and corona, respectively. Due to the lack of low-energy coverage by \nustar{}, we fixed the hydrogen column density at $0.63\times 10^{22}$ cm$^{-2}$ \citep{1990ARA&A..28..215D, 2016A&A...594A.116H}. The continuum emission is characterized by a disk emission of temperature $kT_{in}\sim 0.95$ keV and a power law component with a photon index of $\Gamma\sim 1.58$ and the cutoff energy $E_{cut}\sim 76\kev{}$. The high-energy cutoff is well constrained with the $3-70$ \kev{} \nustar{} spectrum. All the continuum model parameters are listed in Table~\ref{continuum}. Most importantly, the residual of the fit revealed the presence of disc reflection features, a broad Fe K$\alpha$ line around $6-8$ \kev{} and a Compton hump peaking at $\sim 20$ \kev{} (see left panel of Figure~\ref{Fig2}).\\

To describe the continuum emission precisely, we employed a physically-motivated thermal Comptonization model {\tt nthcomp} \citep{1996MNRAS.283..193Z, 1999MNRAS.309..561Z} instead of {\tt cutoffpl}. The {\tt nthcomp} model attempts to simulate the upscattering of photons through the corona from a seed spectrum parameterized either by the inner temperature of the accretion disc or the NS surface/boundary layer. We explored both possibilities in our analysis. We applied the Comptonization model {\tt nthcomp}, setting the photon seed input to a disc blackbody and also to a single-temperature blackbody. Either of these shapes can be selected via the $inp-type$ model parameter. The model {\tt const*TBabs*(diskbb+nthcomp)} with $inp-type=1$ (Model 2) and $0$ (Model 3) describes the continuum emission well with $\chi^2/dof=1637/1506$ and $\chi^2/dof=1633/1505$, respectively. 
The Comptonization model, with input seed photons from the accretion disc and the neutron star surface/boundary layer, statistically describes the continuum emission in a similar manner. The spectral fit with Model 2 and Model 3 reveals a seed photon temperature of $kT_{bb}\sim 0.75$ keV (tied with the $kT_{in}$) for $inp-type=1$ and an upper limit of $kT_{bb}\lesssim 0.90$ (not well constrained by the data) for $inp-type=0$. In the latter case, for $inp-type=0$, the inferred low seed-photon temperature (only an upper limit) makes it difficult to identify their origin. In addition, we note that for both cases $inp-type=1$ and $inp-type=0$, the disc temperatures are comparable within uncertainties, $kT_{in}\sim 0.74$ keV and $kT_{in}\sim 0.81$ keV, respectively. We estimated the electron temperature at $kT_{e}= 22\pm 1 \kev{}$ with photon index $\Gamma= 1.79\pm 0.01$ for $inp-type=1$ and for $inp-type=0$ those values are quite similar (see Table~\ref{continuum}). The electron temperature of the corona, $kT_{e}$, obtained from both scenarios is comparable to the observed high-energy cutoff value obtained from the {\tt cutoffpl} model. The $kT_{e}$ value is also consistent with the high electron temperatures commonly inferred in hard-state neutron star systems \citep{2000ApJ...533..329B, 2006ChJAS...6a.183D, 2017MNRAS.468.3979P}. All the model parameters for the continuum emission are quoted in Table~\ref{continuum}. Moreover, we found that the Comptonized emission dominates the spectrum, contributing around $90\%$ of the total unabsorbed flux.\\

\subsubsection{NICER}
\begin{figure*}
\centering
\includegraphics[width=0.31\textwidth]{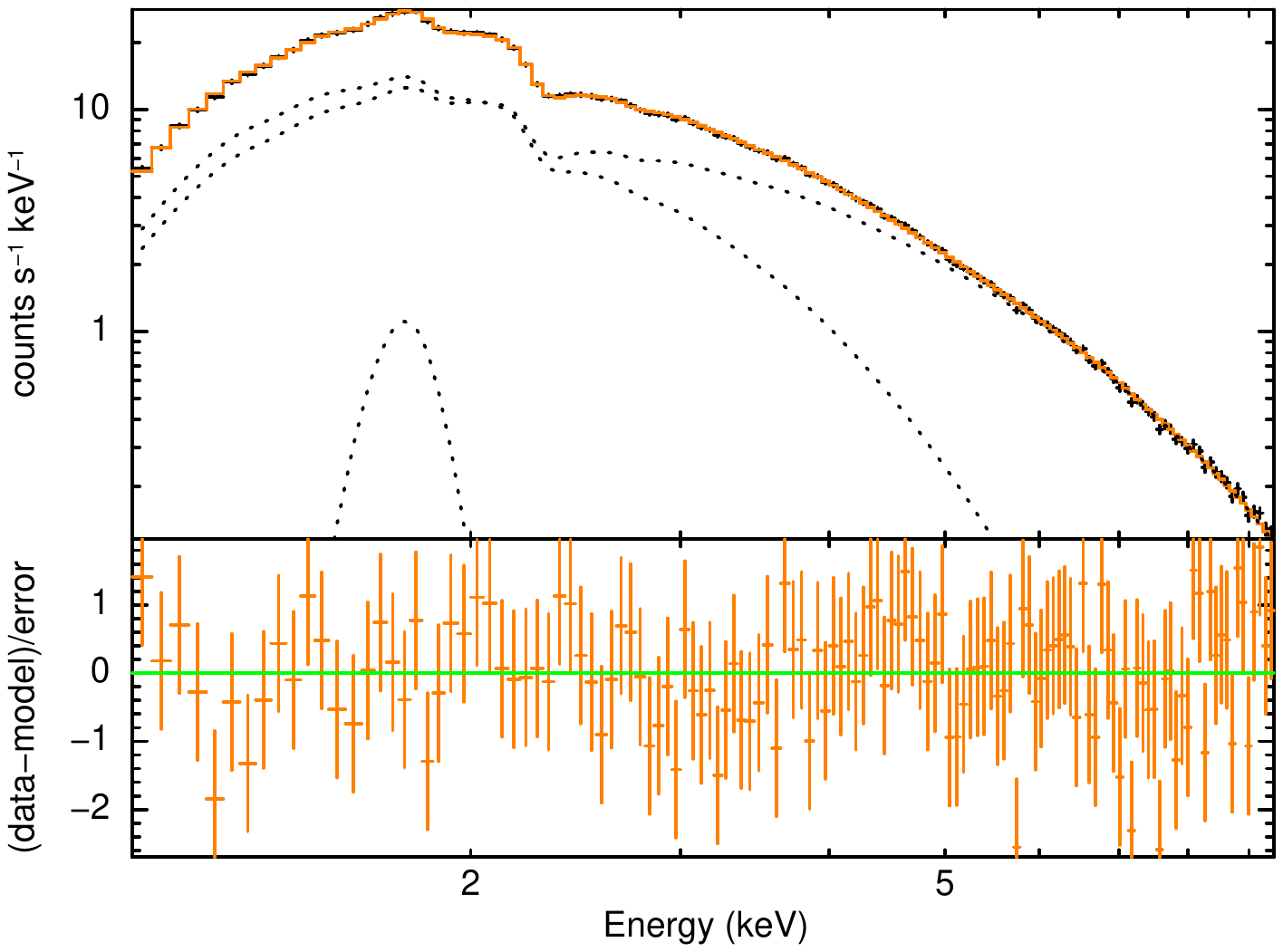}
\includegraphics[width=0.31\textwidth]{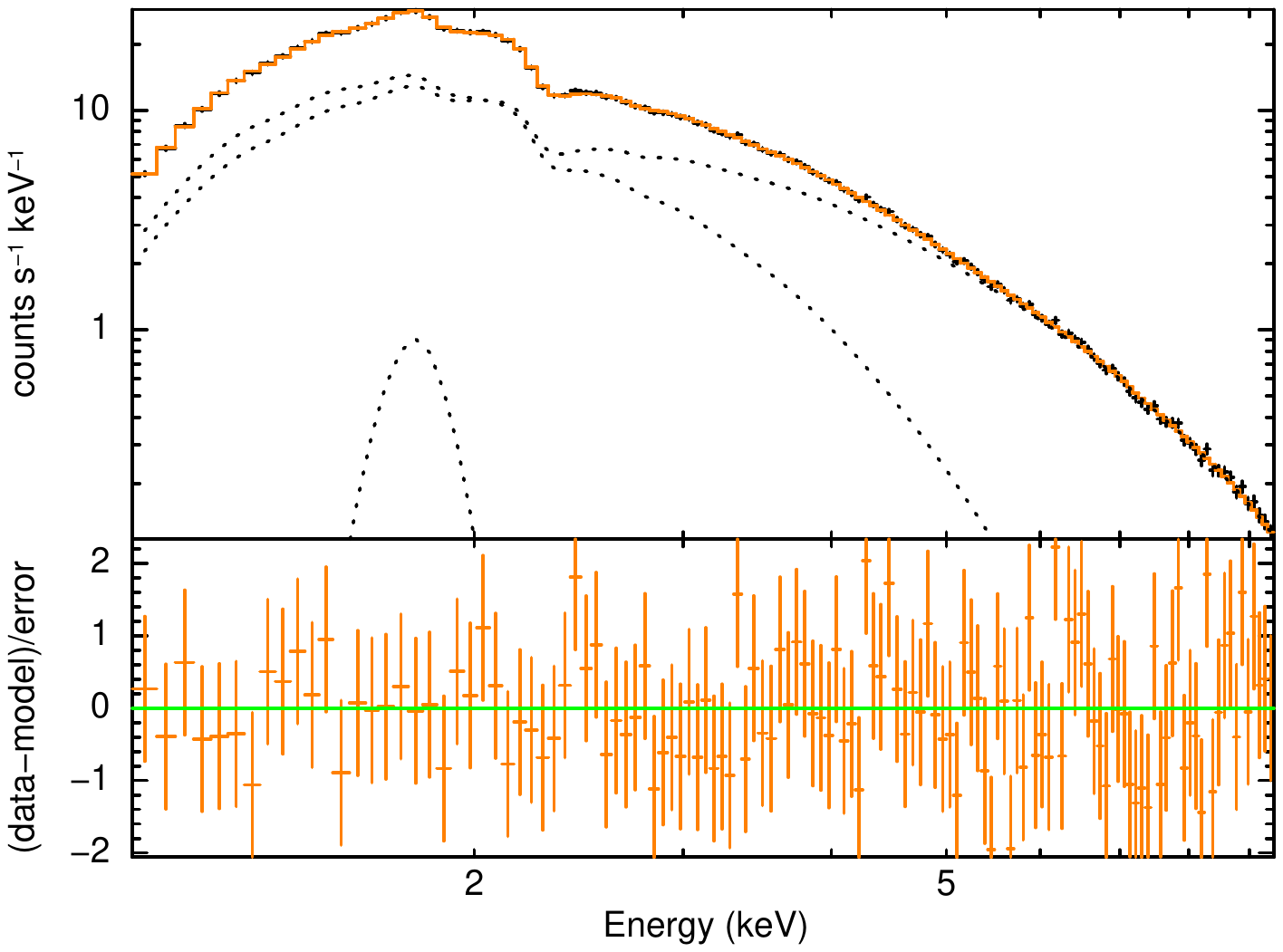}
\includegraphics[width=0.31\textwidth]{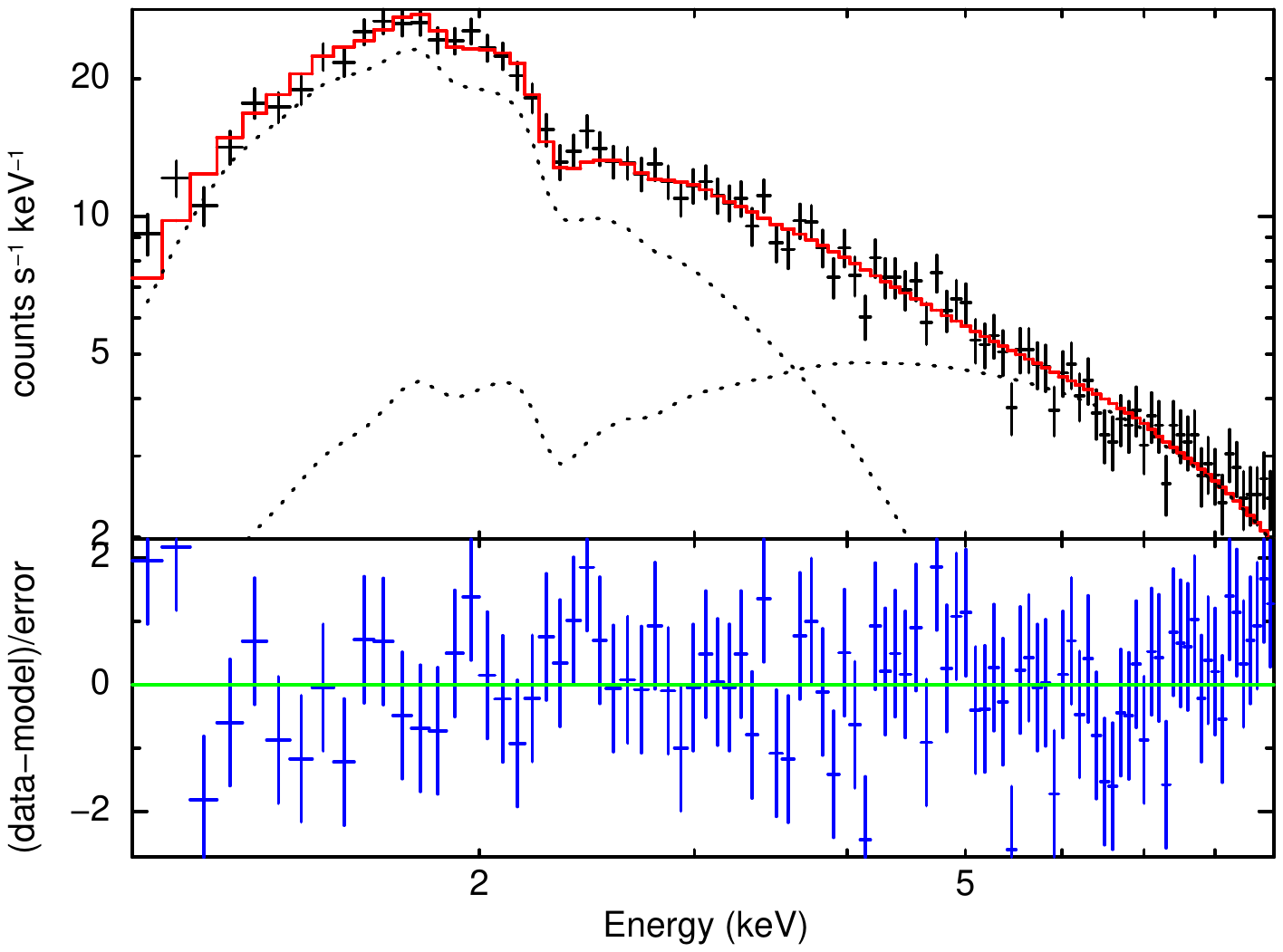}
\caption{ \nicer{}/XTI spectrum of the source 4U~1812-12 in the energy band $1.0-9.5$ keV for obs3 and obs 11 (according to Table~\ref{nicer_obs}) are shown in the left and middle panels. The unfolded spectral data for the continuum model {\tt TBabs*(diskbb+powerlaw+gaussian)} is shown. The right panel shows the spectrum for the flaring part in the energy band $1.0-9.0$ keV, which is modeled with {\tt TBabs*(diskbb+nthcomp)}.  The lower panel of all the plots shows the ratio of the data to the model in units of $\sigma$.}
\label{Fig3}
\end{figure*}

We fit the \nicer{} spectra of this source in the energy band $1.0-9.5$ keV using a model combination of absorbed {\tt diskbb} plus {\tt powerlaw}. For all \nicer{} spectra, we detected an emission line $\sim 1.8$ keV, which is fitted using a narrow {\tt Gaussian} model of fixed width of $0.1$ keV. This emission line is an instrumental artifact known as the Silicon (Si) instrumental edge/feature. The model combination {\tt TBabs*(diskbb+powerlaw+gaussian)}, referred to as Model 4, described all the \nicer{} spectra satisfactorily with an acceptable $\chi^2/dof$ values. We then tried to replace the {\tt powerlaw} model with {\tt nthcomp}, but found it difficult to constrain the {\tt nthcomp} parameter values due to limited data statistics. In particular, the electron temperatures ($kT_{e}$) remained completely unconstrained because of the lack of high-energy coverage by \nicer{}. We found that the best-fitting values of $N_{H}$ vary between $1.40\times 10^{22}$ cm$^{-2}$ and $1.53\times 10^{22}$ cm$^{-2}$ for the \nicer{} observations performed between 2019 and 2021. The inferred $N_{H}$ value is consistent with the \inte{} and \sax{} observations of this source \citep{2006A&A...448..335T}. However, we noticed that the Galactic absorption in the direction of the source is estimated to be $\sim 0.63\times 10^{22}$ cm$^{-2}$ \citep{2016A&A...594A.116H}. A similar discrepancy with the $N_{H}$ value was observed earlier for another ultra-compact binary, MAXI J1957+032 \citep{2022MNRAS.516L..76S}. All the best-fit parameters are quoted in the Table~\ref{nicer_parameter}. The unfolded \nicer{} spectra for two observations are shown in the left and middle panel of Figure~\ref{Fig3} and the variation of the crucial parameters over \nicer{} observation time is shown in the Figure~\ref{Fig5}.\\

We note that one of the \nicer{} observations (obs9) shows regular flaring activity, as shown in the bottom panel of Figure~\ref{Fig1}. For obs9, we generated GTI files for the flaring regions (flares 1, 2, and 3) and extracted spectra for all three regions. As for the persistent emission, we initially modeled the spectra during the flaring phase in the $1.0-9.0$ keV energy band using an absorbed {\tt diskbb} plus {\tt powerlaw} model. However, it failed to describe the spectra, and the {\tt powerlaw} photon index revealed a negative value ($\Gamma\sim -2$), which is unphysical. We then replaced the {\tt powerlaw} component with the physical model {\tt nthcomp}. We tied the seed photon temperature ($kT_{bb}$) to the disk temperature ($kT_{in}$) by accounting for the distribution of the seed photons from the disc. The model {\tt TBabs*(diskbb+nthcomp)} (Model 2) provided an acceptable fit for all the flaring spectra ($\chi^2/dof$ are $87/90$, $114/94$, and $110/90$ for flares 1, 2, and 3, respectively).  However, {\tt powerlaw} $\Gamma$ remained unconstrained in all cases and pegged at the hard limit. We note that a slightly high $\chi^2/dof$ value for flares 2 and 3 is due to the presence of an absorption-edge-like feature $\sim 1.1$ keV. We observed a rise in disc temperature of $\sim 1.12$ keV during the flaring phase compared with the persistent phase. The Comptonization component, which dominates the spectrum (comprising around $80\%$ of the total flux), revealed an electron temperature of around $\sim 19$ keV, consistent with the hard spectral state. The \nicer{} spectrum, along with unfolded spectral models for flare 1, is shown in the right panel of Figure~\ref{Fig3}. All the best-fit parameters are quoted in Table~\ref{nicer_flares}.

\begin{table*}
\centering
\caption{Best-fit parameter values of the \nustar{} spectrum of the source 4U~1812-12 using the Model 5: {\tt const*TBabs*(diskbb + relxill)}.}
\begin{tabular}{lcc}
\hline
Component & Parameter (unit) & Model 5\\
\hline

CONSTANT & FPMB (with FPMA) & $1.005\pm 0.002$ \\

TBabs & $N_{\rm H}$ ($\times 10^{22}$ cm$^{-2}$) & $0.63$ (f) \\

Diskbb & $kT_{\rm in}$ (keV) & $0.75 \pm 0.05$ \\
& Norm & $13^{+4}_{-3}$ \\

RELXILL & $i$ (deg) & $25^{+4}_{-3}$ \\

& $R_{\rm in}$ ($R_{\rm ISCO}$) & $\leq 1.72$ \\

& $\Gamma$ & $1.73 \pm 0.03$ \\

& $\log\xi$ (erg cm s$^{-1}$) & $2.72 \pm 0.08$ \\

& $A_{\rm Fe}$ (solar) & $\leq 0.72$ \\

& $E_{\rm cut}$ (keV) & $130^{+21}_{-18}$ \\

& $refl_{frac} $& $0.15 \pm 0.03$ \\

& Norm ($\times 10^{-2}$) & $(1.15 \pm 0.03)$ \\

& Spin parameter ($a$) & $0.0$(f) \\

CFLUX & $F_{\rm diskbb}$ ($\times 10^{-10}$ erg s$^{-1}$ cm$^{-2}$) & $0.11 \pm 0.01$ \\

& $F_{\rm relxill}$ ($\times 10^{-10}$ erg s$^{-1}$ cm$^{-2}$) & $7.30 \pm 0.02$ \\

& $F_{\rm total}$ ($\times 10^{-10}$ erg s$^{-1}$ cm$^{-2}$) & $7.40 \pm 0.03$ \\

\hline
& $\chi^2_\nu$ (d.o.f) & 1.01 (1501) \\

\hline
\end{tabular}
\label{nustar_parameter}
\end{table*}

\begin{sidewaystable*}[t]
\centering

\caption{Best-fit spectral parameters obtained with the model {\tt TBabs(diskbb+powerlaw+gaussian)}, referred to as Model 4. Errors are quoted at the 90\% confidence level. The Gaussian line width ($\sigma$) was fixed at 0.1 keV.}

\setlength{\tabcolsep}{2.5pt}
\renewcommand{\arraystretch}{1.8}

\begin{tabular}{lccccccccccccc}
\hline

&
TBabs &
\multicolumn{2}{c}{diskbb} &
\multicolumn{2}{c}{powerlaw} &
\multicolumn{2}{c}{gaussian} &
\multicolumn{1}{c}{Flux} &
\multicolumn{2}{c}{$\chi^2_\nu$ (d.o.f.)}&
\\

ObsID &
$N_{\rm H}$ ($10^{22}$ cm$^{-2}$) &
$kT_{\rm in}$ (keV) &
Norm &
$\Gamma$ &
Norm ($10^{-2}$) &
$E_{\rm line}$ (keV) &
Norm ($10^{-4}$) &
{($10^{-10}$ erg cm$^{-2}$ s$^{-1}$)} & 
&
\\

\hline

Obs 3 &
$1.47\pm0.02$ &
$0.76\pm0.02$ &
$24.6^{+2.26}_{-1.98}$ &
$1.39\pm5.0$ &
$3.4\pm0.34$ &
$1.75\pm0.03 $&
$4.0\pm0.98$ & $ 2.76\pm 0.04$&
$0.76 (125)$ \\

Obs 4 &
$1.46\pm0.02$ &
$0.76\pm0.019$ &
$26.5^{+2.30}_{-2.05}$ &
$1.21\pm0.05$ &
$2.4\pm0.25$ &
$1.75\pm0.05$ &
$3.2\pm0.94$&
$ 2.76\pm 0.04 $
& 
$0.77 (124) $\\

Obs 5 &
$1.50\pm0.02$ &
$0.69\pm0.02$ &
$32.3^{+3.26}_{-2.87}$ &
$1.29\pm0.05$ &
$2.3\pm0.21$ &
$1.78\pm0.05$ &
$1.8\pm0.86$ & $2.19\pm 0.01$ &
$0.60 (118)$ \\

Obs 7 &
$1.47\pm0.02$ &
$0.72\pm0.02$ &
$27.7^{+2.64}_{-2.34}$ &
$1.21\pm0.05$ &
$2.0\pm0.20$ &
$1.80\pm0.06$ &
$2.1\pm0.82$ & $2.22\pm 0.01$&

$0.64 (118)$ \\

Obs 8 &
$1.40\pm0.02$ &
$0.82\pm0.02$ &
$22.2^{+1.78}_{-1.63}$ &
$0.82\pm0.07$ &
$1.06\pm0.14$ &
$1.70\pm0.04$ &
$3.9\pm0.94$ & $2.55\pm 0.02$ &
$1.08 (122)$ \\

Obs 10 &
$1.47\pm0.02$ &
$0.72\pm0.02$ &
$27.4^{+2.76}_{-2.43}$ &
$1.22\pm0.05$ &
$2.0\pm0.20$ &
$1.76\pm0.05$ &
$2.1\pm0.85$ & $2.16\pm 0.01$ &
$0.77 (116)$ \\

Obs 11 &
$1.49\pm0.02$ &
$0.74\pm0.02$ &
$28.8^{+2.83}_{-2.47}$ &
$1.40\pm0.05$ &
$3.6\pm0.37$ &
$1.79\pm0.05$ &
$1.4\pm 0.46$ & $2.83\pm 0.04$&
$0.73 (121) $\\

Obs 12 &
$1.50\pm0.02$ &
$0.75\pm0.02$ &
$26.6^{+2.70}_{-2.33}$ &
$1.47\pm0.05$ &
$4.7\pm0.48$ &
$1.8\pm0.06$ &
$3.4\pm1.17$ & $3.19\pm 0.05$ &
$0.63 (122)$ \\

Obs 13 &
$1.45\pm0.02$ &
$0.78\pm0.02$ &
$24.9^{+2.38}_{-2.08}$ &
$1.26\pm0.06$ &
$3.0\pm0.37$ &
$1.76\pm0.04$ &
$1.6\pm0.48$ & $3.07\pm 0.04$&
$0.71 (121) $\\

Obs 14 &
$1.43\pm0.02$ &
$0.81\pm0.02$ &
$25.2^{+2.23}_{-2.00}$ &
$1.01\pm0.06$ &
$2.0\pm0.26$ &
$1.73\pm0.03$ &
$2.1\pm0.48$ &$ 3.23\pm 0.05$&
$0.71 (122)$ \\

Obs 15 &
$1.45\pm0.02$ &
$0.77\pm0.02$ &
$27.3^{+2.81}_{-2.46}$ &
$1.21\pm0.07$ &
$2.6\pm 0.35$ &
$1.75\pm0.03$ &
$1.8\pm0.49$  & $2.93\pm 0.04$ &
$0.81 (119) $\\

Obs 16 &
$1.53\pm0.03$ &
$0.73\pm0.03$ &
$30.2^{+4.21}_{-3.43}$ &
$1.51\pm0.06$ &
$5.3\pm0.64$ &
$1.68\pm0.11$ &
$2.9\pm1.58$& $3.33\pm 0.04$ &
$0.97 (118)$ \\

Obs 17 &
$1.45\pm0.02$ &
$0.80\pm0.03$ &
$25.5^{+2.87}_{-2.52}$ &
$0.93\pm0.09$ &
$1.5\pm0.27$ &
$1.71\pm0.04$ &
$4.1\pm1.18$& $2.77\pm 0.03$&
$0.87 (115)$ \\

Obs 18 &
$1.40\pm0.02$ &
$0.82\pm0.02$ &
$22.2^{+1.78}_{-1.63}$ &
$0.82\pm0.07$ &
$1.1\pm0.14$ &
$1.70\pm0.04$ &
$3.9\pm0.94$& $2.29\pm 0.01$ &
$1.09 (122)$ \\

Obs 19 &
$1.43\pm0.02$ &
$0.80\pm0.03$ &
$24.8^{+3.03}_{-2.62}$ &
$0.95\pm0.10$ &
$1.4\pm0.28$ &
$1.62\pm0.07$ &
$3.9\pm1.31$ & $2.62\pm 0.02$ &
$0.99 (114)$ \\

Obs 20 &
$1.47\pm0.03$ &
$0.76\pm0.03$ &
$24.8^{+3.38}_{-2.78}$ &
$1.37\pm0.09$ &
$3.2\pm0.60$ &
$1.73\pm0.12$ &
$1.7\pm1.34$ & $2.71\pm 0.02$&
$0.77 (112)$ \\

Obs 21 &
$1.49\pm0.03$ &
$0.69\pm0.03$ &
$28.9^{+4.95}_{-3.95}$ &
$1.36\pm0.08$ &
$2.6\pm0.43$ &
$1.78\pm0.08$ &
$2.2\pm1.10$  & $2.13\pm 0.01$ &
$0.86 (107)$ \\

Obs 22 &
$1.52\pm0.04$ &
$0.71\pm0.04$ &
$28.6^{+5.60}_{-4.27}$ &
$1.51\pm0.08$ &
$4.3\pm0.72$ &
$1.73\pm0.08$ &
$2.9\pm1.47$ & $2.69\pm 0.02$ &
$0.73 (110) $\\

Obs 23 &
$1.42\pm0.02$ &
$0.84\pm0.03$ &
$26.4^{+3.21}_{-2.83}$ &
$0.71\pm 0.11$ &
$1.1\pm0.27$ &
$1.79\pm0.05$ &
$3.6\pm1.35$ & $3.36\pm 0.05$&
$0.94 (114)$ \\

Obs 24 &
$1.43\pm0.03$ &
$0.81\pm0.03$ &
$25.4^{+3.66}_{-3.09}$ &
$0.94\pm0.13$ &
$1.5\pm0.42$ &
$1.66\pm0.10$ &
$3.2\pm1.61$ & $2.85\pm 0.02$&
$1.02 (107)$ \\

Obs 26 &
$1.49\pm0.04$ &
$0.77\pm0.04$ &
$27.9^{+4.89}_{-3.88}$ &
$1.26\pm0.12$ &
$2.7\pm0.70$ &
$1.71\pm0.04$ &
$5.3\pm1.73$ &$ 2.80\pm 0.01$ &
$0.81 (104)$ \\

\hline
\label{nicer_parameter}
\end{tabular}

\end{sidewaystable*}

\begin{table*}[htbp]
\centering
\caption{ The best-fit spectral parameters (Model 2) for the flaring phases of one of the \nicer{} observations (Obs ID: 2560010203).}
\renewcommand{\arraystretch}{2.0}

\begin{tabular}{ccccc}
\hline
\textbf{Model Component} & \textbf{Parameter} & \textbf{Flare 1} & \textbf{Flare 2} & \textbf{Flare 3} \\
\hline
TBabs & $N_{\rm H}$ ($10^{22}$ cm$^{-2}$) & $1.18\pm 0.05$ & $1.13\pm 0.04$ & $1.25\pm 0.04$ \\
Diskbb & $kT_{\rm in}$ (keV) & $1.12\pm 0.08$ & $1.08\pm 0.04$ & $0.95^{+0.16}_{-0.04}$ \\
& Norm & $8^{+3}_{-1}$ & $9^{+2}_{-1}$ & $15\pm 2$ \\
nthComp & $\Gamma$ & $1.01^{+0.01}_{pegged}$ & $1.01^{+0.01}_{pegged}$ & $1.01^{+0.01}_{pegged}$ \\
& $kT_e$ (keV) & $19^{+3}_{-5}$& $19\pm 3$ & $18^{+3}_{-1}$ \\
& $kT_{\rm bb}$ (keV) & $=kT_{\rm in}$ & $=kT_{\rm in}$ & $=kT_{\rm in}$ \\
& Norm ($\times10^{-3}$) & $5.57^{+4.16}_{-2.19}$ & $4.38^{+1.99}_{-1.22}$ &  $4.18^{+1.51}_{-1.05}$\\
& Input Type & 1 (fixed) & 1 (fixed) & 1 (fixed) \\
Flux ($1-10$ keV) & $(10^{-9}$ erg cm$^{-2}$ s$^{-1})$ & $1.28\pm 0.03$ & $1.06\pm 0.01$ & $1.13\pm 0.01$ \\
\hline
 & $\chi^2$/dof & $87/90$ & $114/94$ & $110/90$ \\
\hline
\end{tabular}
\label{nicer_flares}
\end{table*}

\subsection{Reflection spectroscopy}
The residuals for the \nustar{} data for all continuum models show an emission line near $\sim 6-7$ \kev{} and a broad excess peaking at around $\sim 20$ \kev{}, which indicate the presence of the Fe K$\alpha$ line complex and the Compton hump, respectively (see left panel of Figure~\ref{Fig2}). These features are known to be produced by the reflection of the hard photons originating from the corona or neutron star surface/boundary layer to the accretion disc. As the continuum emission of the \nustar{} spectrum is well described by the combination of a {\tt diskbb} plus a {\tt cutoffpl} model, we considered a model of reflection off a photoionized accretion disc that is illuminated by a powerlaw emission. For the reflection of a power law, we employed the standard reflection model {\tt relxill}, which provides an illuminating power law with photon index $\Gamma$ and a high-energy cutoff. {\tt relxill} contains the {\tt xillver} model \citep{2010ApJ...718..695G} of reflection off a photoionized disc and {\tt relline} code \citep{2010MNRAS.409.1534D} to account the relativistic effects. In the {\tt relxill} model, we let the model return the primary continuum and the expected reflected spectrum from the disc, while {\tt diskbb} accounts for the temperature and normalization of the accretion disc. \\

\begin{figure*}
\centering
\includegraphics[width=0.4\textwidth, scale=0.40]{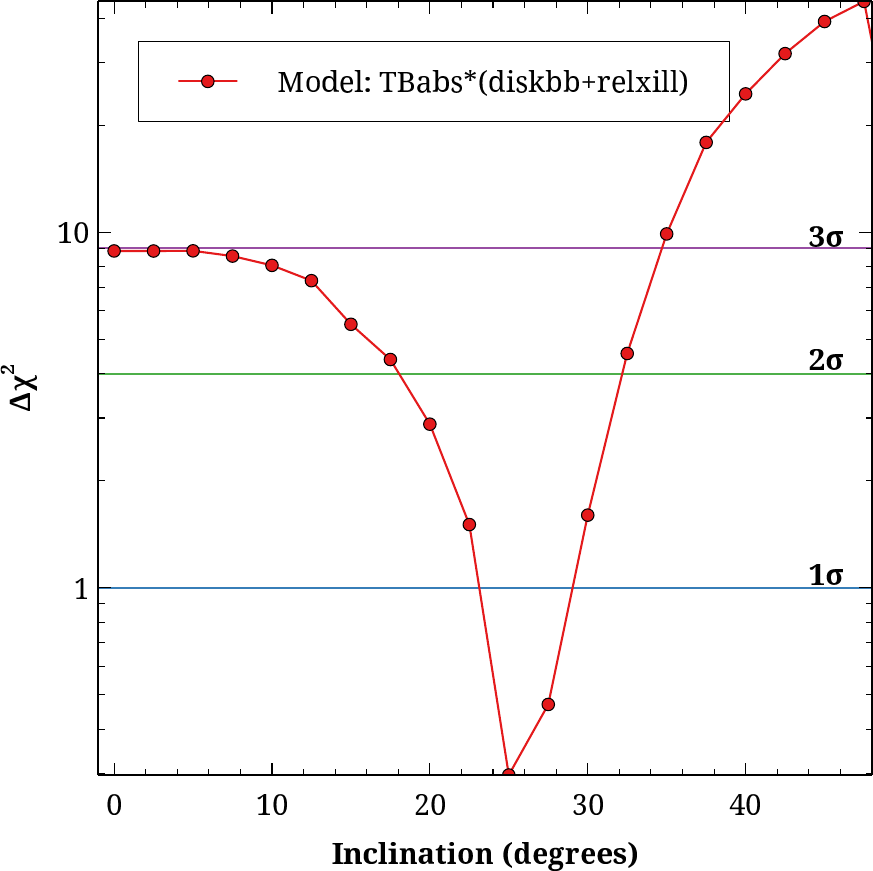}
\includegraphics[width=0.4\textwidth, scale=0.40]{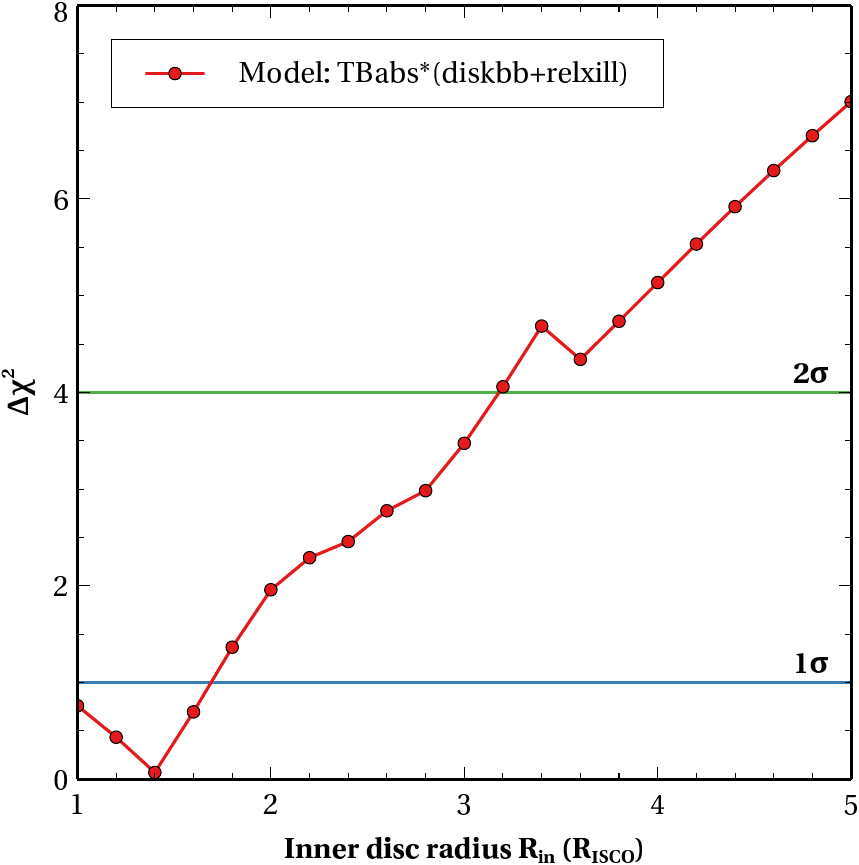}
\caption{It shows the variation of the $\Delta\chi^2$ Surface with the inclination angle and the inner disc radius (in units of $R_{ISCO}$) for the best-fit Model 5. Vertical lines show the $1\sigma$, $2\sigma$, and $3\sigma$ confidence levels.} 
\label{Fig4}
\end{figure*}

\begin{figure}
\centering
\includegraphics[scale=0.32]{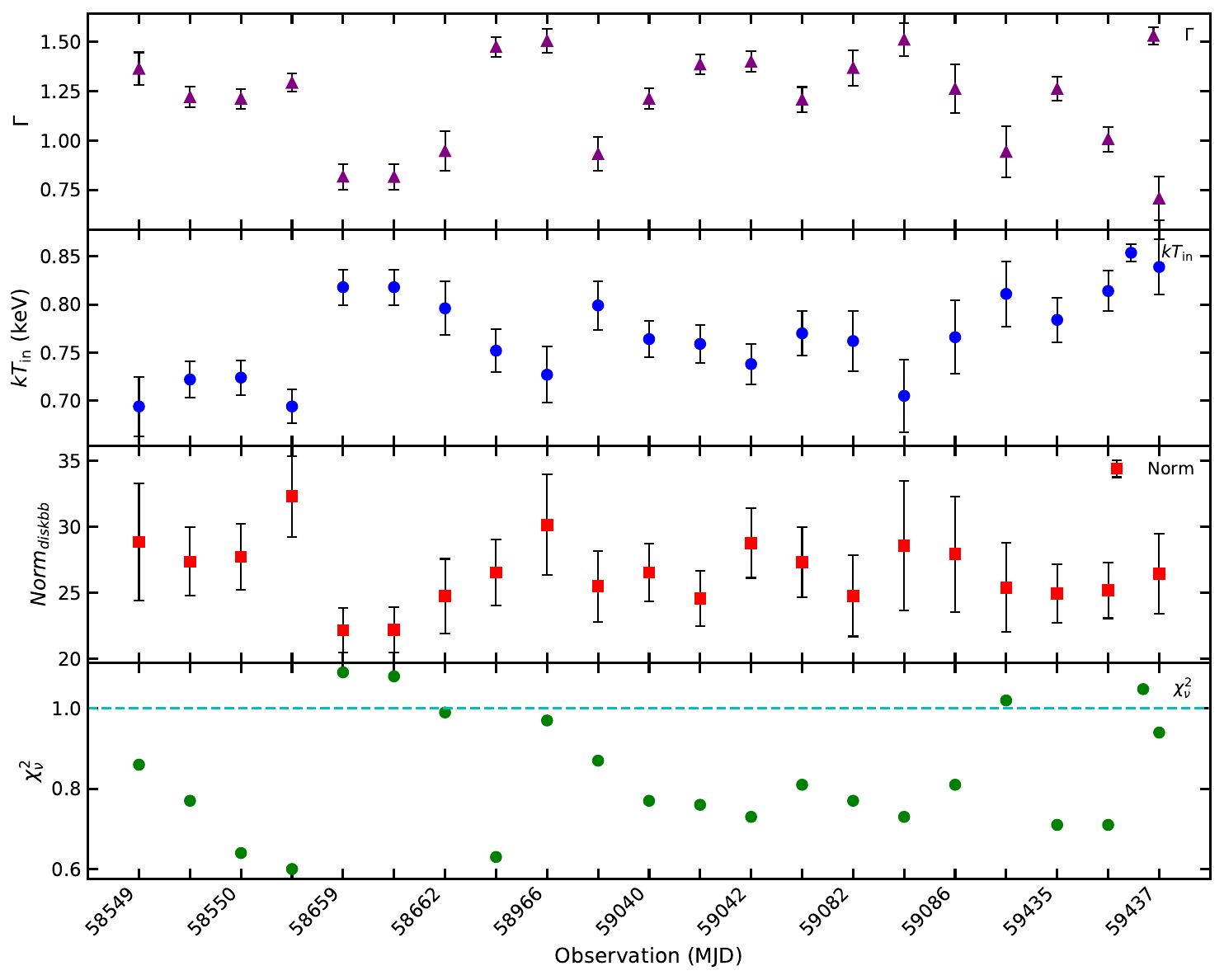}
\caption{This plot shows the variation of the parameters (power law $\Gamma$, disc temperature $kT_{in}$, disc normalization $N_{diskbb}$, and reduced chi-square $\chi^{2}_{\nu}$ values from top to bottom) for the best-fit Model 4 {\tt TBabs*(diskbb+powerlaw+gaussian)} obtained from \nicer{} observations.} 
\label{Fig5}
\end{figure}

We fixed some of the {\tt relxill} model parameters in our analysis. We assumed an unbroken emissivity profile with index $q=3$, consistent with theoretical prediction and observations \citep{2010ApJ...720..205C, 2012MNRAS.424.1284W}. We set the dimensionless spin parameter $a$ to zero, since the source's spin frequency is unknown. For NSs, the spin $a$ typically ranges from $0.0$ to $0.3$, where it only minimally impacts the surrounding metric. We tested that setting $a=0.3$ does not yielded significant changes in either the model parameters or the fit statistics. Other parameters like inner disc radius $R_{in}$, inclination $i$, power law photon index $\Gamma$, disc ionization log$\xi$, iron abundance $A_{Fe}$, reflection fraction $f_{refl}$, and normalization are kept free during fitting. The model {\tt TBabs*(diskbb+relxill)}, referred to as Model 5, revealed a best fit with $\chi^{2}/dof=1518/1501$. The spectral components and ratio of the data to the overall model is shown in the right panel of Figure~\ref{Fig2}. Most of the parameters are well constrained and are listed in Table~\ref{nustar_parameter}. The reflection component implied a $1\sigma$ upper limit of the inner disk radius of $\sim 1.72\;R_{ISCO}$. In addition, the model yielded a low inclination estimate of $\sim 25$ degree and an intermediate disk ionization of log$\xi\sim 2.72$, consistent with the typical range observed in NS LMXBs (log$\xi\sim 2-3$). The upper limit of the iron abundance is comparable to the solar abundance and the reflection fraction is low $\sim 0.15$, indicates that the reflected emission contributes only $\sim 15\%$ of the direct continuum, as expected from the characteristics of the reflection features (see Figure~\ref{Fig2}). The spectrum is significantly harder, yielding a higher cutoff energy of $E_{cut}\sim 130$ \kev{}. The emission is totally dominated by the Componized continuum and its reflection, comprised of $\sim 98\%$ of the total unabsorbed flux, relative to the disc emission. Figure~\ref{Fig4} shows the $\Delta\chi^2$ confidence contours obtained by using the command {\tt steppar} in {\tt XSPEC} for two important parameters ($R_{in}$ and $i$) of the best-fit model {\tt TBabs*(diskbb+relxill)}. In the left panel of Figure~\ref{Fig4}, the sharp minimum indicates that the inclination ($i$) is well constrained by the reflection spectrum. The right panel of Figure~\ref{Fig4} indicates that the disc is consistent with extending close to the neutron star, with no compelling evidence for a strongly truncated disc.\\

We tried to implement another self-consistent relativistic reflection model, {\tt relxillCP}, since our continuum model is also well described by an absorbed {\tt diskbb} plus a Comptonization model, {\tt nthcomp}. However, we obtained an inner disc temperature of $\sim 0.75-0.85$ keV from our fitting with the {\tt nthcomp} model. It prevents us from using the reflection model {\tt relxillCP}, as it has a hard-coded disk temperature of $\sim 0.01$ keV, which is much lower than the disc temperature inferred by the {\tt nthcomp} model.

\section{Discussion}
We present a detailed spectral and timing analysis of the \nicer{} and \nustar{} observations of the source 4U~1812-12, aimed at constraining its accretion geometry. During these observations, the source remained in the so-called hard spectral state, dominated by Comptonized emission. The continuum emission, as characterized by the \nustar{} observation, is well described by the combination of a multi-color disc blackbody component ({\tt diskbb}) and a Comptonized emission from the corona, modeled by either {\tt powerlaw} or {\tt nthcomp}. The \nustar{} spectrum is significantly harder, yielding a higher energy cutoff value ($E_{cut}\sim 76$ \kev{}) and high electron temperature ($kT_{e}\sim 22$ \kev{}) from the continuum modeling with {\tt powerlaw} and {\tt nthcomp}, respectively. The \nicer{} spectra are also well described by a soft thermal disc emission ({\tt diskbb}) and a dominant non-thermal power law ({\tt powerlaw}) emission from the corona. The disc emission is characterized by a temperature of $\sim 0.69-0.84$ \kev{} and a normalization of $\sim 22-32$ from both \nicer{} and \nustar{} observations. The hard component may arise because of the Comptonization of the seed photons originating from the accretion disc, as revealed by the \nustar{} spectrum. Although \nicer{} spectra exhibit a hard spectrum, the emission geometry could not be confirmed due to limited data statistics. Most importantly, the \nustar{} spectrum shows clear evidence of disc reflection, a broad Fe K$\alpha$ emission line in the energy range $6-8$ \kev{} and a Compton hump at $15-30$ \kev{}. The overall $3-70$ \kev{} \nustar{} spectrum is modeled by a soft thermal emission from the disc, a hard Comptonized emission from the corona, and its reflection from the accretion disc. We employed the standard self-consistent relativistic reflection model {\tt relxill} to study the reflection spectrum, which also accounts for the primary X-ray continuum. \nicer{} $1.0-10.0$ \kev{} flux revealed an average value of $\sim 2.81\times 10^{-10}$ ergs cm$^{-2}$ s$^{-1}$. However, during the flaring phases, it has increased to an average value of $\sim 1.16\times 10^{-9}$ ergs cm$^{-2}$ s$^{-1}$. The $3.0-70.0$ \kev{} unabsorbed flux, as revealed by \nustar{} spectrum, is $\sim 7.40\times 10^{-10}$ ergs cm$^{-2}$ s$^{-1}$, which corresponds to a persistent low luminosity of $L=1.49\times 10^{36}$ ergs s$^{-1}$ (assuming a distance of 4.1 kpc). Therefore, we confirm that the source detected in the low-flux state during the \nicer{} and \nustar{} observations, which is consistent with the previous measurements done by different authors \citep{2000A&A...357..527C, 2003A&A...400..643B, 2005ApJ...626.1020M, 2006A&A...448..335T}. Moreover, the low persistent X-ray luminosity points towards the ultracompact nature of this source. \\
 Our spectral modeling from \nicer{} observations aimed to constrain the disc parameters and their variation across the \nicer{} observations. We also wanted to examine the correlation between the disc temperature, $kT_{in}$, and the power law index, $\Gamma$. Figure~\ref{Fig5} shows the variation in disc temperature, disc normalization, and the power-law index over time. We found that the disc temperature exhibits a small variation, maintaining a moderately low value of $\sim 0.69-0.84$ \kev{} and the disc normalization varies within $22-32$ during the \nicer{} observations performed between 2019 and 2021. It indicates that the accretion disc is thermally stable throughout the \nicer{} observations. The power-law photon index, $\Gamma$, exhibits a large variation of $\sim 0.8-1.5$, implying a substantial change in the Comptonized emission. Despite these fluctuations, the consistently low value of $\Gamma$ indicate that the source remains in a hard spectral state. In addition, around the time when the disc temperature increases, the power-law gamma decreases markedly, and vice versa. However, the lack of a clear correlation between $kT_{in}$ and $\Gamma$ suggests that changes in the coronal properties, such as its optical depth, electron temperature or geometry, also play an important role. Various X-ray spectral and timing studies have already noted that the corona geometry changes with the change of the accretion rate \citep{2019Natur.565..198K, 2025ApJ...980..251A}. However, the exact nature of the corona evolution remains unclear. We further extracted and analyzed spectra from the flaring phase of this source and found that the flaring spectra are also dominated by Comptonized emission from a hot corona at $\sim 20$ keV, which accounts for $\sim 80\%$ of the total flux. We further noted from the spectral analysis of the flaring region that during the flaring phase, the disk temperature increased by $\sim 40\% $ compared to the persistent one, reaching a maximum of $\sim 1.12$ keV. Also, an enhancement of $1-10$ keV flux by a factor of $\sim 3$ is observed during the flaring. The flaring behavior in NS LMXBs is actually caused by different physical conditions \citep{2012A&A...546A..35C}. However, previous studies have reported that it is usually triggered by disc instabilities when a massive mass injection travels inward \citep{2018ApJ...867...64C, 2023ApJ...957...27M}. Therefore, the observed increase in disc temperature may be related to the rapid increase in the local mass accretion rate through the inner disk.\\
 
In addition, we performed the PDS analysis using \nicer{} observations after extracting a $0.1$-s binned light curve, which revealed broadband aperiodic variability over $0.01-5$ Hz for most observations. However, one of the \nicer{} observations (obs 4) shows two narrow Lorentzian features at $0.379\pm 0.008$ and $0.724\pm 0.025$ Hz, with Q values of $\sim 5$ and $\sim 9$, respectively. Those have been identified as the fundamental and (sub)harmonic QPO features with fractional rms amplitudes of $2.9\pm 0.6\%$ and $4.1\pm 0.5\%$, respectively. The results obtained from the PDS analysis are in line with those reported in the \rxte{}/PCA PDS analysis of this source by \citet{2003A&A...400..643B}.\\

The \nustar{} spectrum shows clear evidence of disc reflection, a broad Fe K$\alpha$ emission line in the energy range $6-8$ \kev{} and a Compton hump at $15-30$ \kev{}. This is the first detection of the iron K$\alpha$ line and reflection components from the ultra-compact binary 4U~1812-12. However, evidence for iron lines and reflection components has been reported earlier for compact systems such as NGC 6440 X-2 \citep{2010ApJ...714..894H}, MAXI J0911-655 \citep{2017A&A...598A..34S}, and IGR J17062-6143 \citep{2017MNRAS.464..398D}. From the reflection spectrum, we measured a $1\sigma$ upper limit of the inner disk radius of $R_{in}\lesssim 1.72\;R_{ISCO}\sim 10.32\;R_{g}\sim 22$ km and a low inclination of $\sim 25$ degrees. It indicates that the accretion disc remains close to the neutron star or associated with a little disc truncation during the \nustar{} observation. Such a low inclination is also consistent with the absence of eclipses or absorption dips that are typically observed in high-inclination X-ray binaries \citep{1987A&A...178..137F}. In addition, the reflection model also estimated an intermediate disk ionization of log$\xi\sim 2.72$, consistent with the typical range observed in NS LMXBs (log$\xi\sim 2-3$) and an iron abundance less/comparable to the solar value. The reflecting disc exhibits only a weak reflection fraction ($\sim 15\%$), suggesting that although the disc extends close to the NS, only a small fraction of the coronal emission is intercepted and reflected. This is consistent with NS LMXBs in the hard state, where Comptonized emission dominates over the thermal disc component \citep{2015MNRAS.449.2794D, 2017ApJ...836..140L, 2019MNRAS.490.2300M, 2020MNRAS.494.3177M}. \\

We estimated a bolometric flux of $F_{bol}=(9.50\pm 0.02)\times 10^{-10}$ ergs cm$^{-2}$ s$^{-1}$ by extrapolating the \nustar{} best fit over the $0.01-100$ \kev{} range. This value corresponds to the bolometric luminosity of $L_{bol}\sim 1.90\times 10^{36}$ ergs s$^{-1}$, assuming a source distance of 4.1 kpc. We used this $L_{bol}$ value to estimate the mass accretion rate, which was found to be $\dot{m}\sim 1.62\times 10^{-10}\:M_{\odot}$ yr$^{-1}$, consistent with an average mass accretion rate of $\sim 3\times 10^{-10}\:M_{\odot}$ yr$^{-1}$ \citep{2000A&A...357..527C}. We can estimate the magnetic field strength of the source using Equation (1) in \citet{2009ApJ...694L..21C}, assuming that the disc is truncated by the magnetic field. Using geometrical and efficiency parameters from \citet{2009ApJ...694L..21C}, we found $B\lesssim 2.54\times 10^{8}$ G for $\eta=0.1$, $f_{ang}=1$, and $k_{A}=0.5$ \citep{2009MNRAS.400..492I}. This estimates is comparable to the typical values observed for NS LMXBs \citep{2015MNRAS.452.3994M}.

\section{Data availability}
This research has made use of data obtained from the HEASARC, provided by NASA's Goddard Space Flight Center. The \nustar{} and \nicer{} observational data sets used in this work are in the public domain and is available on NASA's website https://heasarc.gsfc.nasa.gov. 
 
\section{Acknowledgements}
This research has made use of data and/or software provided by the High Energy Astrophysics Science Archive Research Centre (HEASARC). This research also has made use of the \nustar{} data analysis software ({\tt NuSTARDAS}) jointly developed by the ASI Space Science Data Center (SSDC, Italy) and the California Institute of Technology (Caltech, USA). ASM would like to thank Inter-University Centre for Astronomy and Astrophysics (IUCAA) for their facilities extended to him under their Visiting Associate Programme.

\bibliography{4u1812}{}
\bibliographystyle{aasjournalv7}

\end{document}